\documentclass[10pt,twocolumn]{article}
\usepackage{latex8}
\usepackage{times}
\usepackage{graphicx}
\usepackage{amsmath}
\usepackage{color}
\usepackage{url}

\newcommand{\bfA}{{\mathbf A}}

\newcommand{\bfa}{{\mathbf a}}

\newcommand{\bfd}{{\mathbf d}}

\newcommand{\bff}{{\mathbf f}}
\newcommand{\bfg}{{\mathbf g}}

\newcommand{\Cp}{C_{\mathrm{p}}}
\newcommand{\CT}{C_{\mathrm{T}}}

\newtheorem{alg}{Algorithm}
\begin{document}

\title{Reducing the noise effects in Logan graphic analysis for PET receptor measurements\thanks{
This work was supported by grants from the state of Arizona, the NIH (R01 AG031581, R01 MH057899 and P30 AG19610) and the NSF (DMS 0652833 and DMS 0513214).
}}
\author{Hongbin~Guo\thanks{Corresponding author. Email: hb\_guo@asu.edu,Tel: 480-965-8002,   Fax: 480-965-4160.}\ \  \thanks{Department of Mathematics and Statistics, Arizona State University, Tempe, AZ  85287-1804. },\ \  Kewei~Chen$^{\S}$,\ \ Rosemary~A~Renaut$^{\ddag}$\ \ and\ \   Eric~M~Reiman\thanks{Banner Alzheimer Institute and Banner Good
Samaritan Positron Emission Tomography Center, Phoenix, AZ 85006.}\\
}

\maketitle

\begin{abstract}
 Logan's graphical analysis (LGA) is a widely-used approach for quantification of biochemical and physiological processes  from Positron emission tomography (PET) image data. A well-noted problem associated with the LGA method is the bias in the estimated parameters. We recently systematically evaluated the bias associated with the linear model approximation and developed an alternative to minimize the bias due to model error.  In this study, we examined the noise structure in the equations defining linear quantification methods, including LGA. The noise structure conflicts with the conditions given by the Gauss-Markov theorem for the least squares (LS) solution to generate the best linear unbiased estimator. By carefully taking care of the data error structure, we propose to use structured total least squares (STLS) to obtain the solution using a one-dimensional optimization problem.  Simulations of PET data for [11C] benzothiazole-aniline (Pittsburgh Compound-B [PIB]) show that the proposed method significantly reduces the bias.  We conclude that the bias associated with noise is primarily due to the unusual structure of he correlated noise and it can be reduced with the proposed STLS method.
\end{abstract}

%-------------------------------------------------------------------------
\Section{Introduction}

Graphical analysis (GA)  is a routine tool for quantitative imaging with PET in various clinical and physiological studies. The first GA method is the commonly usedd Patlak method that was introduced by Patlak, \cite{Patlak83,Patlak85}, for irreversible tracers. Logan extended this method for reversible tracers, \cite{Logan90}. So far a set of GA methods have been developed for both reversible and irreversible systems, for plasma input and reference model, and for calculation of uptake rate, distribution of volume (DV) and DV ratio (DVR) or binding potential (BP). 

A well-noted problem with the use of GA methods, particularly for reversible system, is the bias in the estimated parameters, \cite{slifstein2000, ichise2002str}.  To reduce the bias,  Logan {\it et al} suggest to deal with noise by smoothing the data, \cite{logan2001a}, and Varga {\it et al} proposed perpendicular least squares, \cite{varga2002mod}, which is exactly the total least squares (TLS) method, \cite{Golub73}. Ichise {\it et al}, \cite{ichise2002str}, rearranged the equation to a multilinear equation to decrease the noise, \cite{ichise2002str}. Ogden used a nonlinear likelihood estimation, \cite{ogden2003est}. The  bias associated with GA approaches, we believe, has three possible sources. Published work primarily dealt with  the bias related to the random noise, one of  the three sources. The other two sources are the numerical quadrature error and  an approximation of the underlying compartmental model. We recently systematically evaluated the bias associated with the model approximation and developed an alternative model for  minimizing the bias caused by model error, \cite{Guo_moderr_appl08}.  In this study, we investigate the noise effects in parameter estimation differently. In contrast to the linear least squares (LLS) or ordinary TLS algorithms, careful examination of the data error structure leads to our proposal to develop  a structured total least squares (STLS) approach to estimate the parameters. Simulation shows that the bias due to noise is greatly reduced by the STLS method. 

The rest of the paper is organized as follows: The new approach, STLS,  is introduced in Section~\ref{sec:method}. The simulation study is described in  Section~\ref{sec:valid} and results  reported in Section~\ref{sec:results}. Issues relevant to the proposed approach are discussed in Section \ref{sec:dis}.  Conclusions are presented in Section~\ref{sec:conc}.

\section{Methods} \label{sec:method}
\subsection{Logan's method and alternative linear methods}
Assume the equilibrium is reached at some time point $t'$, after which linear equations associated with corresponding linear methods are assumed valid.  Logan's GA (LGA) quantification method for reversible radiotracers with known plasma input function, i.e. plasma concentration of the unmetabolized tracer,  is based on the following equation
\begin{equation}\label{eq:Logan}
\mathrm{ LGA:}\hspace{0.5cm}\frac{\int_0^t \CT(s){\it ds}}{\CT(t)}\approx V \frac{\int_0^t\Cp(s){\it ds}}{\CT(t)}+b,
\end{equation}
where $\CT(t)$ is the measured ``tissue time activity curve" (TTAC), $\Cp(t)$ is  the input function, and $V$ represents for the volume distribution (DV). The equation approximately reflects the tracer behaviors over the equilibrium period.  Ichise {\it et al}  revised (\ref{eq:Logan}) to a 
 multilinear equation as follows , \cite{ichise2002str},
\begin{equation}\label{eq:Multi-lin}
\mathrm{ MA1:}\hspace{0.1cm} 
\CT(t) \approx -\frac{\mathrm{V}}{b} \int_0^t\Cp(\tau){\rm d}\tau+\frac{1}{b}\int_0^t \CT(\tau){\rm d}\tau.
\end{equation}
Both LGA and MA1 are solved by LLS. In the ordinary least squares method the independent variables are assumed to be noise-free. But Varga {\it et al}, \cite{varga2002mod},  noted that noise appears in both the independent and dependent variables and proposed to use TLS for equation (\ref{eq:MA0}), let us call it MA0, which is also the root equation for LGA and MA1:
\begin{equation}\label{eq:MA0}
\mathrm{ MA0:}\hspace{0.1cm} \int_0^t \CT(\tau){\rm d} \tau \approx \mathrm{DV} \int_0^t\Cp(\tau){\rm d}\tau+b\CT(t).
\end{equation}
The TLS solution is obtained from the right singular vector corresponding to the smallest singular value of the matrix formed by columns $\int_0^t \CT(\tau){\rm d} \tau$, $\int_0^t\Cp(\tau){\rm d}\tau$ and $\CT(t)$. A complete  introduction and analysis of basic algorithms for TLS is presented in \cite{Huffel91}.  %We can think the data smoothing method proposed by  Logan {\it et al} as a pre-processing step,\cite{logan2001a}, which can be combined to any of these linear methods for the purpose of bias reduction. 

\subsection{Method development}
 We denote the durations and central time of the scanning frames by $\Delta_i$ and  $t_i, i=1,\cdots,n$ and assume $t'=t_q$. Thus $t_i, i=q,\cdots,n$ fall in the equilibrium period. Because $\CT(t_i)=\int_{t_i-\Delta_i/2}^{t_i+\Delta_i/2} \CT(s){\it ds}/\Delta_i$ the integral $IC_i=\int_0^{t_i} \CT(s){\it ds}=\sum_{j=1}^{i-1}\CT(t_j)\Delta_j+\CT(t_i)\Delta_i/2$. With these notations the discretized multilinear equation of (\ref{eq:MA0}) can be written as:

\begin{equation*}
\left( \begin{array}{ll}
IP_q, & \CT(t_q)\\
IP_{q+1},   & \CT(t_{q+1})\\
\vdots & \vdots  \\
IP_n , & \CT(t_n)
\end{array}
\right)
  \left( \begin{array}{l}
V\\
b
 \end{array}
 \right)
 \approx
   \left( \begin{array}{l}
IC_q \\
IC_{q+1}\\
  \vdots\\
IC_n
 \end{array}
 \right),
\end{equation*}
where $IP_i=\int_0^{t_i} \Cp(s){\it ds}$. In matrix notation this is
\begin{equation}\label{eq:dis-mlin}
 \bfA   \left( \begin{array}{l}
V\\
b
 \end{array}
 \right)
\approx
\bfd. 
\end{equation}

To simplify the discussion we assume 1) the equilibrium is truly reached after $t'$, thus the discussed linear models do not have model error; and 2) the noise in $\Cp(t)$ can be ignored as compared with the noise in  $\CT(t)$. The  Gauss-Markov theorem tells us that the LLS estimator is the best linear unbiased estimator assuming that no noise in the independent variables and the entries of the noise vector are normally distributed i.i.d. variables with zero mean and common variance. Because both $\CT(t)$ and $\int_0^{t}\CT(s){\it ds}$ contain noise and the noise is correlated, none of Gauss-Markov theorem's conditions are satisfied for the data in linear equations (\ref{eq:Logan}), (\ref{eq:Multi-lin}) and (\ref{eq:MA0}). This explains the bias of these linear methods. On the other hand, the TLS solution is equivalent to the maximum likelihood  solution when the noise in independent and dependent variables are i.i.d multivariate normally distributed with zero mean and common covariance, \cite{Huffel91}. The noise in the linear equations do not satisfy this condition either. Thus TLS does not produce an unbiased parameter either, as noticed in \cite{ichise2002str}.

Let $\CT(t_i)+f_i$  be the true radioactivity at time $t_i$, i.e. $-f_i$ is the measurement error in  $\CT(t_i)$, assumed to be  approximately normal distribution with zero mean and variance  $var(f_i)=\sigma_{i}^2$.   Thus $\gamma=\sum_1^{q-1}f_i\Delta_i $ has variance  $\sum_1^{q-1}(\sigma_{i}\Delta_i)^2$. By incorporating the noise in the approximated equation (\ref{eq:dis-mlin}) we obtain the following exact equation, which reflects the structure of the noise,
\begin{equation}\label{eq:STLS}
\bfA [V,b]^T+b\bff=\bfd+L\bff+\gamma {\bf 1},
\end{equation}
where $\bff=[f_q,\cdots,f_n]^T,{\bf 1}=[1,\cdots,1]^T$, and $L$ is a lower triangular matrix:
$$L=\left(\begin{array}{llll}
      \Delta_q/2,& 0 & \cdots & 0 \\
      \Delta_q ,& \Delta_{q+1}/2 & \ddots & 0 \\
      \vdots & \vdots & \ddots & \vdots\\
      \Delta_q ,& \Delta_{q+1} & \cdots & \Delta_n/2 
    \end{array}\right ).
$$
We propose to estimate $V$ by solving an optimization problem as follows:

\begin{eqnarray}
\mathrm{ STLS:}&&   {\rm min}\quad w\gamma^2+\bff^T E \bff \label{opt:STLS}\\
&&  {\rm subject \  to} \quad   (\ref{eq:STLS}) \nonumber
\end{eqnarray}
where $E={\mathrm diag} (1/\sigma_q^2,\cdots,1/\sigma_n^2)$ and $w=1/(\sum_1^{q-1}(\sigma_{i}\Delta_i)^2)$. (\ref{opt:STLS}) is a STLS problem and equivalent to maximum likelihood (ML) model, as demonstrated  in section \ref{sec:dis}. The $n-q+4$ unknowns are $V,b,\gamma$ and $f_i, i=q,\cdots, n$. For example, if $3$ frames fall in the equilibrium period, i.e. $n-q+1=3$, we have $6$ unknowns.
If we fix the variable $b$ the optimization problem (\ref{opt:STLS}) becomes a quadratic programming problem and can be simplified to a LLS problem as follows. Let us use $\bfa_1$ and $\bfa_2$ to denote the two columns of matrix $A$, i.e. $A=[\bfa_1, \bfa_2]$.  Equation (\ref{eq:STLS}) can be rewritten as 
\begin{equation*}
(L-bI)\bff=[\bfa_1, -{\bf 1}][V,\gamma]^T+ \bfa_2 b-\bfd.
\end{equation*}
Let $B=(L-bI)^{-1}[\bfa_1, -{\bf 1}]$ and $\bfg=(L-bI)^{-1}( \bfa_2 b-\bfd)$ the objective function becomes
$$F(b, V, \gamma)=\left( B\left( \begin{array}{l}
V\\ 
\gamma
 \end{array}
 \right)+\bfg
\right )^TE 
\left( B\left( \begin{array}{l}
V\\ 
\gamma
 \end{array}
 \right)+\bfg
\right )+w\gamma^2.
$$
By setting the first order derivatives of $F(b, V, \gamma)$ with respect to $V$ and $\gamma$  to zero we obtain the following equation
\begin{equation} \label{eq:LS_Vgamma}
 (B^TEB+wJ)\left( \begin{array}{l}
V\\ 
\gamma
 \end{array}
 \right)=-B^TE\bfg,
\end{equation}
where matrix 
$$J=\left( \begin{array}{cc}
0&0\\ 
0&1
 \end{array}\right).$$
Denoting the solution of (\ref{eq:LS_Vgamma}) by $V(b)$ and $\gamma(b)$ we define function $G(b)=F(b,V(b), \gamma(b))$. Based on the above analysis we design a numerical algorithm for solving the STLS problem (\ref{opt:STLS}) by an one dimensional minimization as follows
  
  \begin{alg} \label{alg:STLS}
Given $b\in [\beta_1, \beta_2]$, \\
\begin{enumerate}
\item Solve\quad\quad  ${\rm min} \quad G(b), \ \ for \ \ b\in [\beta_1, \beta_2]$.
\item For the solution of the above minimization, $b^*$, calculate corresponding $V$ by (\ref{eq:LS_Vgamma}).
\end{enumerate}
\end{alg}

\section{Simulation study } \label{sec:valid}
The simulated data are adopted from published clinical data, \cite{Price05, Yaqub08PIB}. Specifically, arterial input function and rate constants of two tissue reversible compartmental model for PIB tracer are used. Eleven regions , ROI {\bf 1}, to ROI {\bf 11},  of normal controls (NC) and  Alzheimer's Disease (AD) diagnosed subjects are tested. Details of the data are described in \cite{Guo_moderr_appl08}.

Given the decay corrected input function and the kinetic parameters we generate corresponding unperturbed TTACs, $\CT^*(t)$. The frame durations, total $240$ minutes scanning, are set to be, given in minutes,  $4\times 0.25$, $8\times 0.5$, $9 \times 1$, $2\times 3$, $8\times 5$ and $18\times 10$.
 Frames falling in 120 to 240 minutes, i.e. frame $37$ to $49$, are chosen as equilibrium frames.
 We set $t'=120$ so that the equilibrium is approximately attained. The only exception is the ROI {\bf 6}, which is far from equilibrium even after $200$ minutes. For the noise-free  decay-corrected concentration TTAC, $\CT^*(t)$, Gaussian  noise at each time point $t_i$, $G(0,\sigma(\CT^*(t))$,  is modeled using the approach in \cite{logan2001a,varga2002mod, ichise2002str} as follows
\begin{equation}\label{var_y}
\sigma(\CT^*(t_i))=Sc\sqrt{ \frac{\CT^*(t_i) e^{\lambda t_i}}{60\Delta t_i}},
\end{equation}
here  $\lambda$ is the tracer decay constant ($0.034$ for $^{11}C$) and $Sc$ is a scale factor, which is set to 0.5 and 1 in our simulations. The units for $\CT^*(t_i)$ and $t_i$ are  kBq/ml and minutes respectively and $\Delta t_i$ is multiplicated by $60$ to measure the frame duration in seconds.  1000 random sample sets are tested for each noise level, Sc=0.5 and Sc=1.

\section{Results}\label{sec:results}
 
\begin{figure}[p]
\includegraphics[scale=.45]{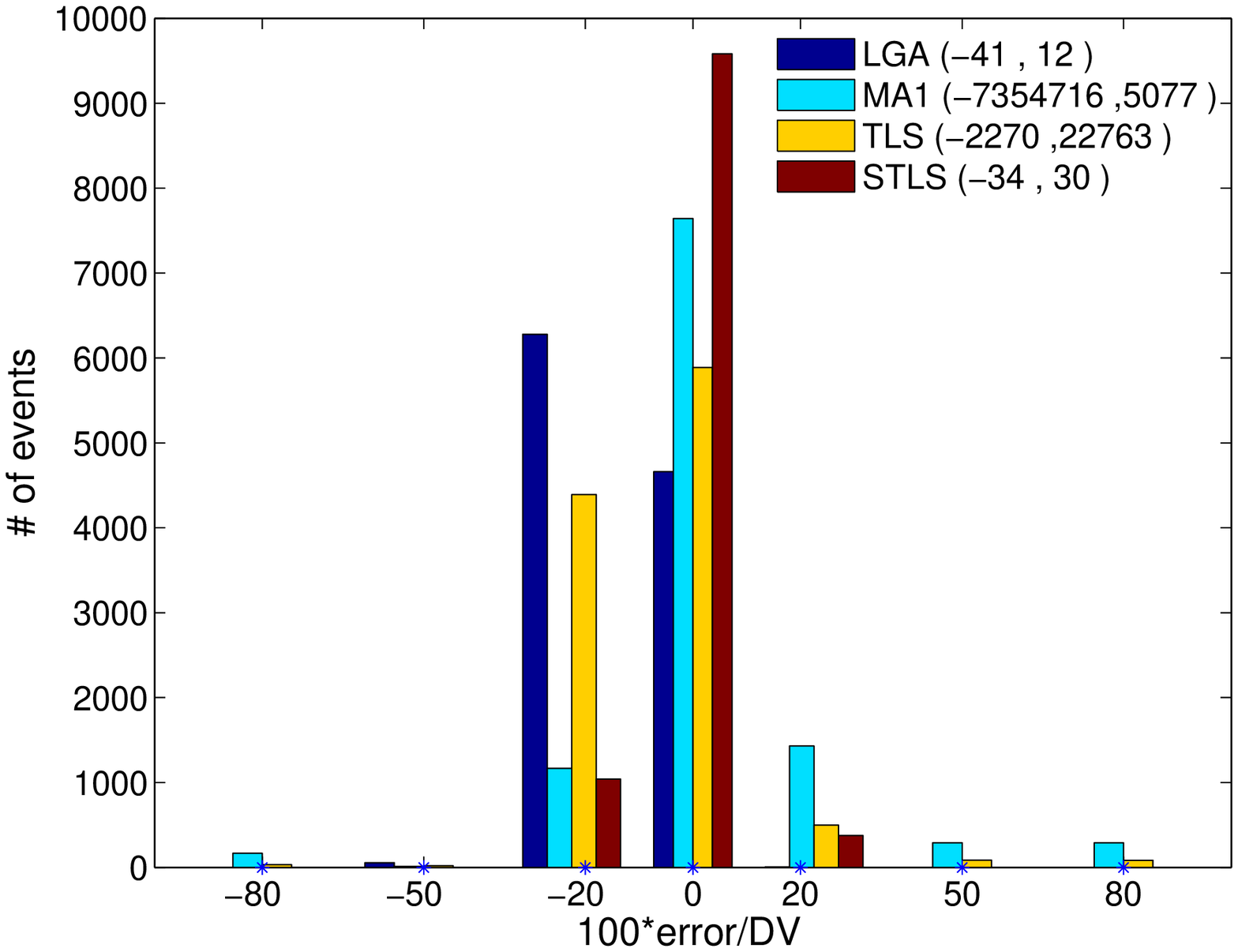}
\includegraphics[scale=.45]{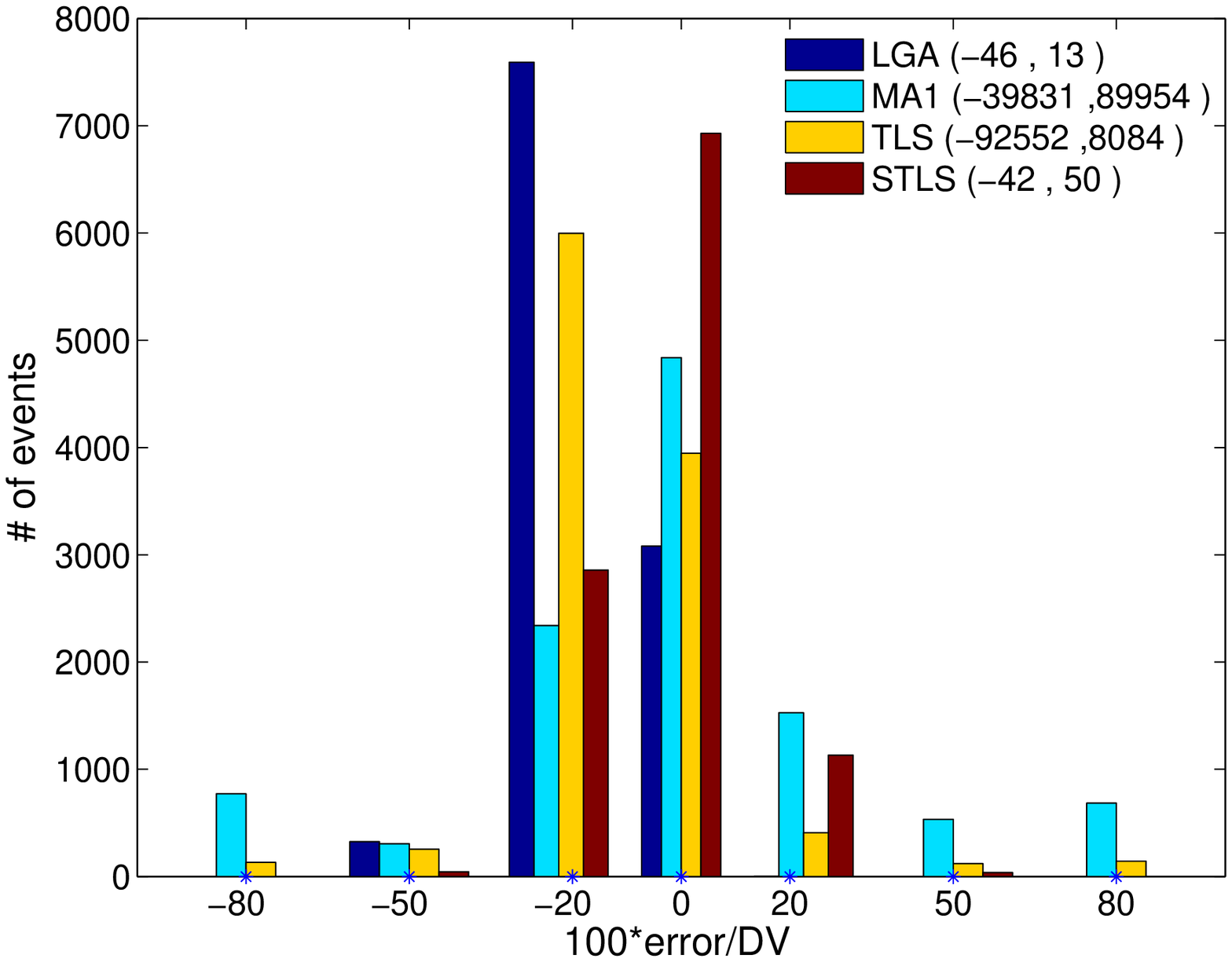}
\caption{Histograms for normalized error (in percentage),$100(\mathrm{V}_{\mathrm{est}}-DV)/DV$, of the results for all eleven ROIs and four methods, LGA, MA1, TLS and STLS. The upper and lower figures are corresponding to noise scale $Sc=0.5$ and $1$ respectively. The error ranges are presented in the legends. The errors are assigned to $7$ bins with centers specified at $-80\%, -50\%, -20\%, 0\%, 20\%, 50\%$ and $80\%$.}
\label{fig:DV-hist}
\end{figure}

We present histograms for the  percentage relative error of the bias, $100(\mathrm{V}_{\mathrm{est}}-DV)/DV$, in Figure~\ref{fig:DV-hist} 
with the range of the percentage error for each method indicated in the legend. The upper figure is for noise scale $Sc=0.5$  while the lower figure is for $Sc=1$. It is clear that the variances for the results of MA1 and TLS and the bias of the results of LGA are too large. STLS outperforms all these three methods.  There are some situations, in which the relative error is less than $-100\%$; in other words, the calculated DVs are negative. This occurs $133$ and $29$ times over all $11000$ random tests for MA1 and TLS at noise scale $Sc=0.5$ while $509$ and $95$ occurrences are observed for MA1 and TLS for noise scale $Sc=1$. LGA and STLS do not produce any negative DV. The reason for the negative DV for MA1 is discussed in  \cite{Guo_moderr_appl08}.   In the simulations, the average CPU time, in seconds, per TTAC  were  $4e-4$, $2e-4$, $3e-4$ and $9.4e-3$, for LGA, MA1, TLS and STLS, respectively.

\begin{figure}[t]
\includegraphics[scale=.45]{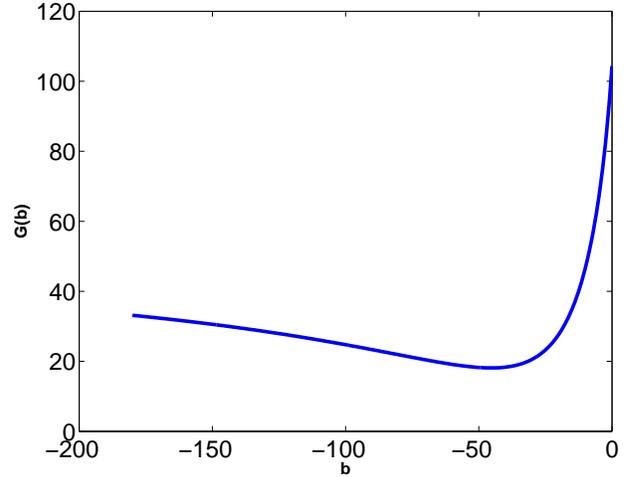}
\caption{Function G(b) for a representative case.}
\label{fig:Gb_curve}
\end{figure}

\section{Discussion} \label{sec:dis}
The proposed STLS is equivalent to the ML solution with the Gaussian density assumption for noise $f_i$. Because the variances for $f_i$ and $\gamma$ are $\sigma_i^2$ and $\Sigma^2=\sum_1^{q-1}(\sigma_{i}\Delta_i)^2$ respectively, the ML problem can be formated as follows
\begin{eqnarray*}
&&   {\rm max}\quad e^{\gamma^2/\Sigma^2}e^{\sum_{i=q}^n f_i^2/\sigma_i^2}\\
&&  {\rm subject \  to} \quad   (\ref{eq:STLS}).
\end{eqnarray*}
The equivalence is easily shown by taking the logarithm of the objective function.

In our simulation the exact variances are used. In practice, the variance of each frame can be estimated, \cite{Carson93var,Pajevic98var}. In this work we assume that the linear simplification for MA0 does not have significant model error, which is the reseason we use a long scan duration $240$ minutes. If the model error does exist, this is the case for practical situations, we need to correct the  model error,  \cite{Guo_moderr_appl08}, and perform further research to reduce the noise effects based on the model error corrected model. At last, in our simulations the range for $b$ is given by $[-150, -10]$ for all cases. A representative curve for function $G(b)$ is illustrated in Figure \ref{fig:Gb_curve}, which is a convex curve. If a case dependent good range of $b$ can be estimated the performace of the STLS algorithm can be further improved. 
\section{Conclusions} \label{sec:conc}
In this article we proposed to reduce the noise effects of linear PET quantification by a structured total least squares noise model and developed an efficient numerical algorithm for its solution. We validated our findings through simulations with clinical derived PIB-PET data.  Simulation
results demonstrate that the STLS algorithm significantly reduces the bias caused by noise in PET data as compared with LGA, MA1 and TLS.

\end{document}